# COVID-19: Open source model for rapid reduction of R to below 1 in high $R_0$ scenarios

Baker, M.R.,a, Hawthorne, E.L.,b, Rogge, J.R.,a
*a Complex Systems Research Division, Medichain Ltd. Euston London, UK*
*b HVS Image Software Ltd. Bicester, UK*

**Abstract**

We present an open source model that allows quantitative prediction of the effects of testing on the rate of spread of COVID-19 described by R, the reproduction number, and on the degree of quarantine, isolation and lockdown required to limit it. The paper uses the model to quantify the outcomes of different test types and regimes, and to identify strategies and tests that can reduce the rate of spread and R value by a factor of between 1.67 and 33.3, reducing it to between 60% and 3% of the initial value.

The model is designed to be simple, transparent and robust and can be run in an Excel spreadsheet, modified easily and shared with or used by stakeholders. Anyone with their own data, national, regional, local or specific (such as for surge testing, border control, specific use cases such as educational establishments etc) can add in their own data and make robust predictions of the level of control of R, and the levels of testing, quarantine and isolation required to achieve it, in different scenarios.

Importantly, they can model the effects of different testing programs in order to choose the program that gives their preferred balance between minimising spread of the disease and minimising isolations, according to initial infection levels, initial R value, the scale of testing possible and other considerations.

Thus the model can be used to reduce transmission while balancing restrictions and freedoms - either in conjunction with vaccination, quarantine and personal protective equipment (PPE) or with any combination or lack of these measures.

We provide a simplified version of the model as an Excel sheet, giving others the ability to adapt, modify, validate, and develop the model and make it available for use in public health for decision making, as well as in open discussion and debate.

The model highlights the value of high sensitivity, medium specificity CLDC testing to bring R below 1 in situations where $R_0$, the basic reproduction number of a variant, is so high that use of the medium sensitivity, high specificity PCR and lateral flow testing, in combination with other measures, fails to do so.

**Introduction**

*Background*

To date there have been 276,436,619 confirmed cases of COVID-19, which is caused by the virus SARS-CoV-2, including 5,374,744 deaths (1), with a significant proportion of survivors experiencing long term effects on their health (2). In addition there have been impacts on other health (such as cancer (3) and mental health (4)) and the global economy, for example an estimated fall in GDP of £216bn in the UK in 2020 (5).

New variants of the virus emerge frequently and those that have evolutionary advantages in the circumstances they emerge in take over as the dominant strains (6). Therefore, new variants of concern have grown exponentially despite measures that had been controlling the previously dominant variant (7, 8, 9, 10).

In populations that have high levels of immunity through prior infection or vaccination, new variants that are highly transmissible and also evade that immunity will have an evolutionary advantage (11, 12) and are likely to become widespread.

A further challenge is that SARS-CoV-2 is most infectious 24-48 hours before symptoms occur and at symptom onset (13). This both causes rapid spread and means there is no evolutionary pressure for new variants not to cause serious disease and death - if a host becomes seriously ill or dies this does not take the virus out of circulation as the host has already transmitted the virus to others earlier, when feeling well and going about their life.

Measures have therefore been put in place in many countries (14) to limit contact between apparently healthy people, in order to attempt to control the reproduction number, R, the number of people infected on average by each one infected person.

At an effective R, $R_e$, of 1, case numbers are steady. If $R_e$ is below 1, case numbers are shrinking and if $R_e$ stayed below 1 everywhere the virus would be eradicated. If $R_e$ is above 1, case numbers are growing exponentially (15). The aim of measures put in place is to reduce $R_e$ by reducing the number of people with whom infectious people come into contact, including those who are infectious but do not yet know it.

*Effectiveness of measures used against different variants of concern*

R for the ancestral variant was reduced in many countries by national lockdowns (16). Test-trace-isolate or test-trace-isolate-support systems then had limited success in controlling the ancestral variant without lockdowns by identifying contacts of known cases and asking them to self-isolate in case they were infectious (17). For example, according to UK government records of case numbers, R fell to below 1 under its test-trace-isolate (TTI) system in June to August 2020(18) (see Figure 1).

With the Alpha variant identified in the UK in September 2020, $R_e$ increased and second and third lockdowns were required, creating a double peaked wave in case numbers in October 2020 to April 2021(18) (see Figure1). Vaccination starting in Spring 2021 allowed lifting of the third lockdown, with $R_e$ for Alpha at just below 1, as seen in gradually falling case numbers from April 2021 (see Figure1):

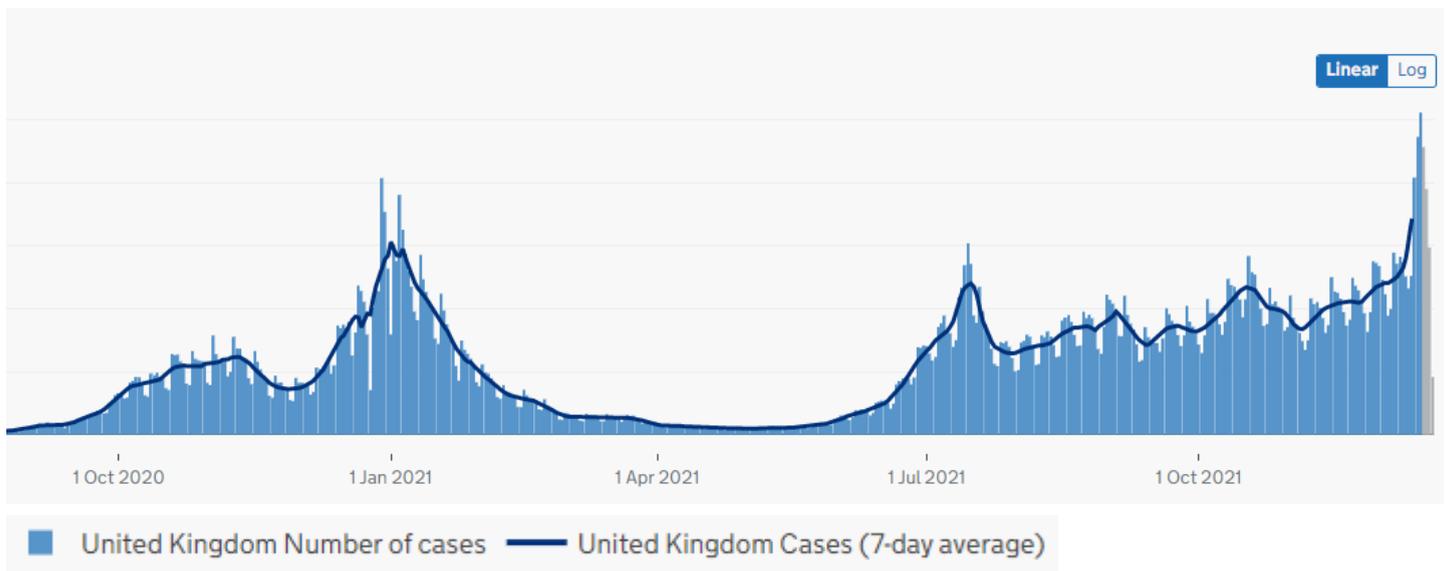

**Figure 1: Cases by specimen date (18)**

By June 2021 the Delta variant had become dominant (19) with $R_e$ again above 1 despite the measures in place, as seen in the exponential rise in case numbers in June to July 2021 (18) (see Figure 1). A combination of TTI, light restrictions and increasing vaccination curtailed growth and vaccination kept serious illness and death down to a level where health services could cope, as seen in the relatively flat numbers of patients in hospital per day from August to December 2021 (see figure 2) (20).

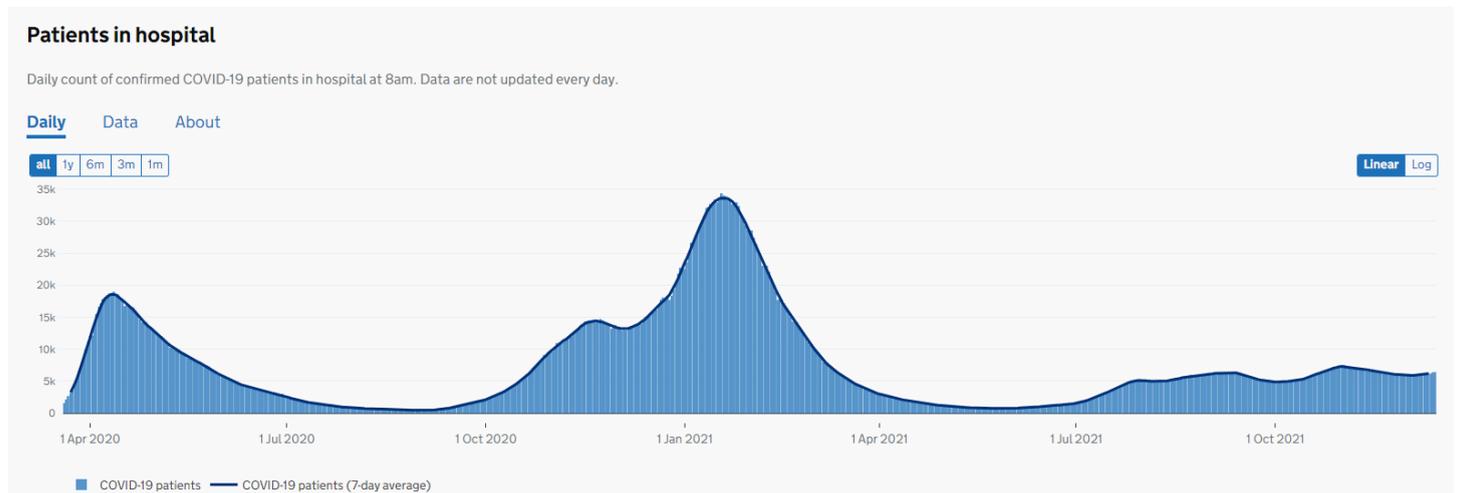

Figure 2: Patients in hospital per day in England (20)

The Omicron variant of concern announced in late November 2021 has been growing exponentially despite the measures already in place against Delta. At the time of writing $R_e$ for Delta in the UK is approximately 1, and $R_e$ for Omicron is estimated to be between 3 and 5 (21). Omicron case numbers have a doubling rate of around 2 days, prompting an emergency vaccination booster program (22) and scientists' advice for tighter social restrictions (23).

The tests currently in use to detect infection (lateral flow and PCR) are estimated to have contributed a 10% to 28% reduction in R via TTI (24); they are of limited value particularly as they have little ability to detect the virus prior to the onset of symptoms, which is part of the most infectious period (13, 17). New variants that thrive even in populations with increased immunity from prior infection or vaccination are therefore likely to spread rapidly despite use of these test types.

### Objective

Our objective here is to provide a simple to use tool that allows anyone such as government advisors and policy makers to assess the health, social and economic impacts of different testing strategies in different circumstances. In particular it allows the assessment of the impact of high sensitivity, medium specificity testing vs medium sensitivity, high specificity testing on reducing R, to avoid the need for stringent social restrictions and lockdowns, including in situations where highly transmissible variants of concern would otherwise grow exponentially despite the use of measures that have been able to limit the spread of previous variants.

By seeing the widely differing impacts of tests with different sensitivities and specificities on the ability to identify infectious individuals and on the proportion of the population required to isolate in different scenarios, government advisors and policy makers could make informed decisions about the type of testing to be carried out, depending on factors such as current R value and the availability of different types of test.

The aim is to allow improved mitigation of Omicron, control of any variant in populations where vaccine supply is short or vaccine hesitancy high, and control of future highly transmissible variants of concern, including those which are even more able to evade immunity from vaccination or prior infection and where the disease needs to be held off while new vaccines are developed, manufactured, tested and administered.

In low R situations, high specificity testing could help control R while minimising the proportion of the population needing to isolate, or high sensitivity testing could be used to bring R well below 1 and eradicate the virus. In high R situations, high sensitivity testing could be used to reduce R much more effectively than high specificity testing, while avoiding the need for mass restrictions or lockdowns.

**Results**

We present results of modelling the effect on the reproduction number, R, of tests with different sensitivities and specificities when used in a test-trace-isolate (TTI) system, with one-off or repeat testing. For each test the model shows the numbers of individuals required to isolate to achieve the reduction and the number allowed to circulate, including how many in each group are infectious and how many are susceptible.

Importantly, the widely used PCR and lateral flow tests, where real world sensitivity is relatively low at approximately 40 - 70% (25-28) but specificity is high at 95 - 99.95%, minimise the number of non-infected individuals required to isolate, but result in significant proportions of infectious individuals circulating, and a relatively high R reduction multiplier. This means that if the initial R level is relatively high the disease is not controlled by these test types, as is seen in real world transmission rates with high R variants such as Omicron (10).

The novel CLDC test, where specificity is relatively low but sensitivity is high (29, 30), minimises the number of infectious individuals circulating and has an R reduction multiplier around 10 times smaller with one or two days of testing (39). This means that even if the initial R level is relatively high the disease is still controlled with CLDC testing.

The model presented here is simplified to a snapshot of the Infectious and Susceptible members of a population at a point in time. Any members of the wider population who are 'Removed' due to prior infection, effective vaccination or death are therefore not relevant to the simplified model, nor are movements from the Susceptible group to the Infectious (or vice versa) which may occur over time.

A more complex version of the model could also show the effect of different frequencies of repeating the test protocol, in order to determine the optimum frequency for maintaining the reduction in R without excessive numbers of individuals having to isolate. The effect of early detection by CLDC (29, 30) could also be modelled in a more complex version to assess the impact of detecting the virus before the other test types are able to, to prevent spread in the particularly infectious period before symptoms occur (13).

*Input parameters:*

Sensitivities and specificities for PCR and lateral flow tests were taken from peer reviewed studies where real world test sensitivities and specificities were determined for tests in widespread use in the UK (25-28). Sensitivity and specificity for CLDC testing is taken from studies undertaken to validate findings of the initial peer reviewed CLDC study (29, 30), with the most conservative (lowest) sensitivity level being used in the model.

A testing population size of 100,000 is used for easy on-the-fly conversion to percentages, with 30% of those tested being infectious. 30% represents a very high but not necessarily unlikely level of infection for a highly transmissible variant (31, 32), and is sufficiently high to allow comparison of the numbers of individuals in each category after testing.

*Output variables:*

The key output variables are the R reduction multiplier; the effective R, $R_e$, after TTI using the stated test type; the number of susceptible individuals isolating as a result of false positive test results; the number of infectious individuals circulating as a result of false negative test results.

*Protocol:*

The protocol to be followed is as follows:

Of all individuals tested on day 1, those who receive a positive result are told to isolate, and those who receive a negative result may circulate.

All those who received a positive result on day 1 are retested on day 2. Those who received a negative result on day 1 are not retested and continue circulating.

Of those retested on day 2, those who receive a positive result again are told to continue isolating, and those who now receive a negative result may now circulate.

This process may continue for n days.

*Outcomes:*

Figures 3 to 6 show the outcomes when n is 2.

An Excel sheet of the model is available for use subject to licence (39). This shows the effect on R of each test type when n is 1, 2, 3, 4 and 5, along with the numbers of individuals required to isolate for the recommended period (e.g. 10 days) (33) to achieve the reduction, and the number allowed to circulate, including how many in each group are infectious and how many are susceptible.

Of particular interest are the R level and whether it is above or below 1, the number of susceptible individuals isolating for the recommended period as a result of false positive test results (which has social, economic and other health impacts), and the number of infectious individuals circulating as a result of false negative test results (which needs to be minimised to reduce transmission).

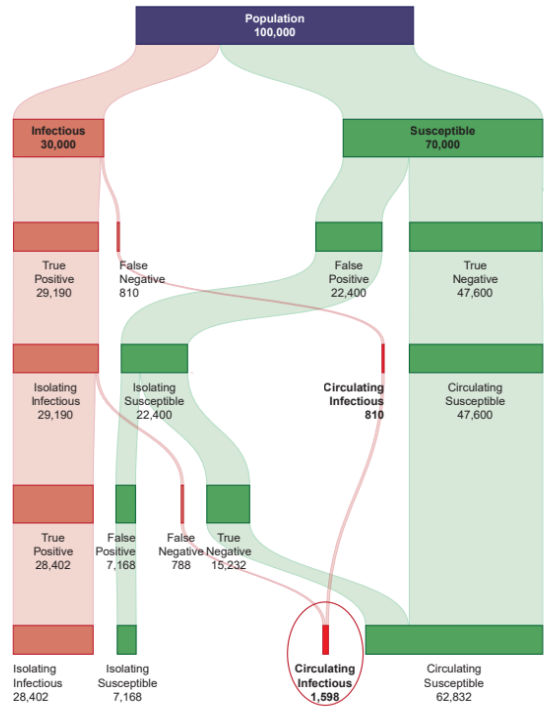

Figure 3

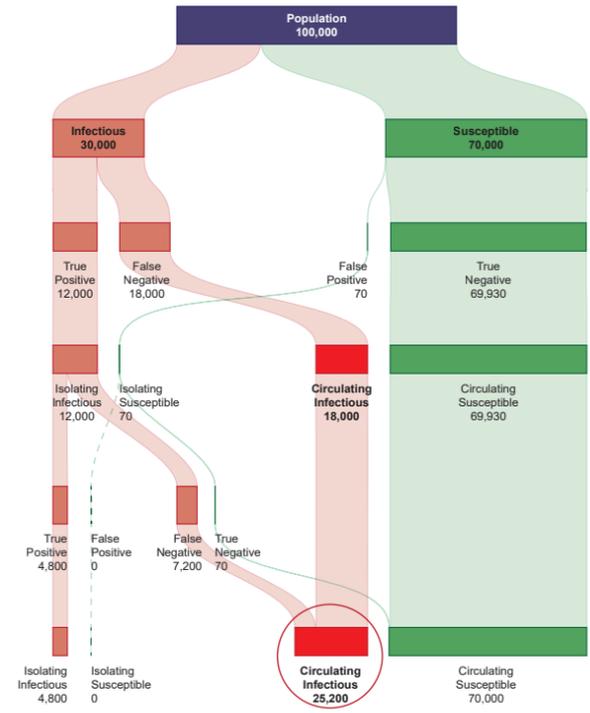

Figure 4

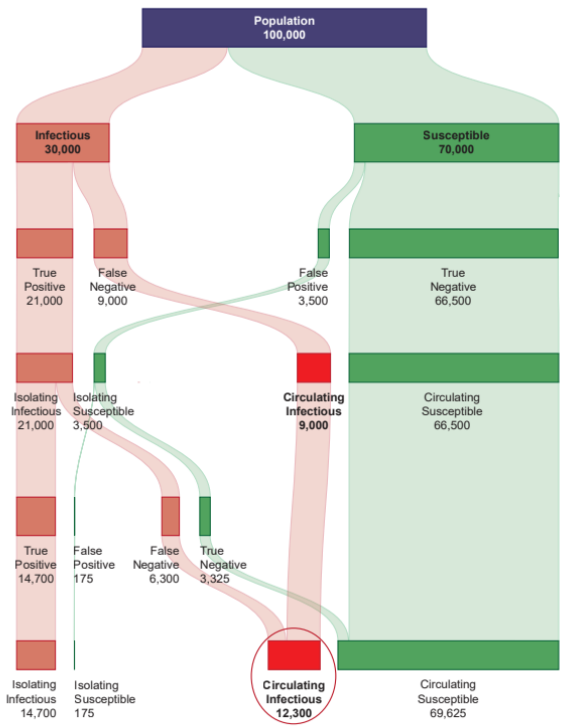

Figure 5

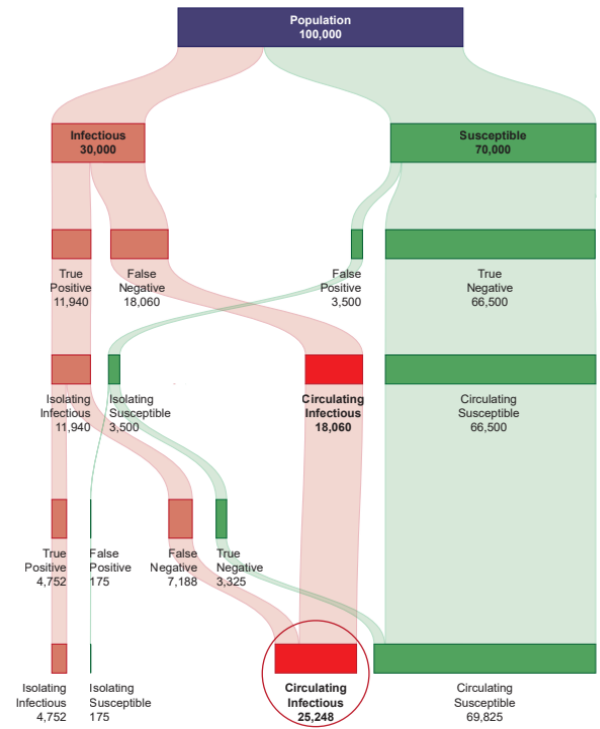

Figure 6

## Discussion

In this study we have modelled the impact of different testing strategies for use in controlling the spread of the SARS-CoV-2 virus, including the use of different types of test with different sensitivities and specificities. The model allows decision makers and their advisors to look at the direct impact of different testing strategies on the reproduction number, R, the key indicator of whether an epidemic or pandemic is growing or shrinking. No mathematical or scientific knowledge is needed for someone to see the relative effects of using different test types with different sensitivities and specificities in test-trace-isolate (TTI), surge testing, mass testing or other approaches. Entering the sensitivity and specificity of any test under consideration shows, for one-off or repeat testing, the number of infectious individuals allowed to circulate, the number of susceptible people required to isolate, and the reduction of the effective reproduction number, $R_e$.

### Understanding Sensitivity and Specificity and their Impact on R:

The implications of test sensitivity and specificity are key to use of testing in virus control. Sensitivity is the ability of a test to correctly identify patients who DO have the disease, and specificity is the ability of a test to correctly identify people who do NOT have the disease (34). They are separate measures of a test's effectiveness and both have impacts that must be taken into account in determining who should isolate in order to reduce R.

If a test's sensitivity is 70%, then for every 100 infectious people tested 70 infectious people are given a correct positive result (true positives) and 30 infectious people are given incorrect negative results (false negatives). If those with positive results self-isolate and those with negative results are allowed to carry on their lives as normal, 30% of the infectious people are circulating in the population, allowing spread of the disease at 30% of the unrestricted rate.

If a test's specificity is 70%, then for every 100 uninfected people tested, 70 of them are given correct negative results (true negatives) and 30 of them are given incorrect positive results (false positives). If those with positive results self-isolate and only those with negative results are allowed to carry on their lives as normal, 70% of the uninfected people will be able to carry on as normal, while 30% of them have to isolate.

Thus in a situation where testing is used to limit spread of COVID-19, by requiring those with positive results to self-isolate and allowing only those with negative results to mix with each other, tests that have specificities close to 100% have the advantage that only a small proportion of uninfected people are asked to self isolate.

The downside is that if these tests have sensitivities much lower than 100%, they result in a significant proportion of the infectious people being allowed to mix freely in the wider population. This is the case with PCR and lateral flow tests where real world sensitivity is 40-70% (25-28).

If the basic R of dominant and emerging variants is low enough, R can be brought below 1 even if a significant proportion of the infected people are still circulating due to false positive test results, i.e. an effective R of 1 or less can be achieved even if sensitivity is significantly lower than 100%.

This was achieved in the UK in April to May 2021, when following national lockdown, measures including vaccination and test-trace-isolate with high specificity, medium sensitivity tests kept the effective R at approximately 1 (reduced from an estimated basic R of the Alpha variant of 3.4 (35)), as seen in the relatively flat case numbers in figure 1 for that period (18).

If however the basic R of dominant or emerging variants is higher, and the same proportion of infected people are circulating due to the same proportion of false negative test results, the same percentage reduction in R will now result in an effective R level that is greater than 1, and case numbers will grow exponentially.

This was the case with the Alpha variant in the UK with TTI with high specificity, medium sensitivity tests in place but *before* vaccination started, with the need for lockdowns in 5 November to 2 December 2020 and 6 January to 8 March 2021 (36), as can be seen in the rising and falling of case numbers in October 2020 to January 2021 in figure 1 (18).

It is also the case with the Omicron variant with TTI using high specificity, medium sensitivity tests in place alongside a high level of vaccination (37), as can be seen in the steep increase in case numbers at the far right of figure 1 (18). Although $R_e$ for Delta had been fluctuating around 1 in the prior period (July to November) with these measures in place, Omicron's basic R is sufficiently high that its $R_e$ with the same measures in place is estimated at 4 - 5 (21).

*Summary*

In variants with relatively low $R_0$, high specificity, medium sensitivity testing alongside mild measures can control R and minimise the isolation of non-infected. In those with relatively high $R_0$, high specificity, medium sensitivity testing doesn't control R unless used in conjunction with stringent measures. In this case the advantage of high specificity tests is outweighed by their disadvantage of insufficient sensitivity - too many infectious individuals receive negative results (false negatives) and go on to infect others.

In the absence of 100% sensitive 100 % specific testing there is a trade-off between those with false positive results isolating unnecessarily and those with false negative results circulating and spreading infection.

The balance point where R is kept at a target level varies depending on what R would be without TTI in place. The model reported here allows this balance point to be found, in order for isolation of non-infected individuals to be no higher than necessary while controlling R and so avoiding escalation to the point where the majority of the population have to isolate to bring it under control.

For example, the model shows an R reduction multiplier of 0.05 achieved with 20.4% of the population isolating for 1 day and 7.17% of the population isolating after a day 2 re-test, contrasting with 81% of adults staying at home in the UK lockdown of 23 March to June 2020 (38).

The ability to find this balance point would be of particular value in situations where serious illness and death are not prevented by vaccination, including where transmission is so high that even a low proportion of serious cases represents a high absolute number, where vaccination levels are low due to hesitancy or lack of availability, or where a new variant has a high degree of vaccine escape.

Extension of the model may include, for example, modelling over a period of time as infection statuses change and the infection level is iteratively reduced by the reduction to R; modelling the effect of detection by CLDC prior to the infectious period (29, 30); adding the ability to predict the number of hospitalizations and lives saved by different testing strategies, the number of person days of lockdown avoided and the impact on essential and non essential services; modelling economic impact and cost benefit analyses using the numbers of person days of isolation.

Use of this model in conjunction with new testing methods has the potential to reduce R, reduce the number of individuals needing to isolate, and make it possible to apply measures earlier and in a more nuanced way to substantially reduce the impact of future waves and mitigate the adverse health, social and economic impacts of COVID-19.

## Methods

The effective R level after TTI, the number of susceptible individuals isolating and the number of infectious individuals circulating are calculated as follows, where $R_e$ is the effective reproduction number achieved after n days of testing, $R_0$ is the reproduction number on day 0, Se is the sensitivity of the test used, $IsS_n$ is the number of susceptible (non-infected) individuals isolating on the nth day, $S_0$ is the initial number of susceptible individuals, Sp is the specificity of the test used, $CI_n$ is the number of infectious individuals circulating on the nth day and $I_0$ is the initial number of infectious individuals:

$$R_e = R_o \times (1 - Se^n)$$

$$IsS_n = S_0 \times (1 - Sp)^n$$

$$CI_n = I_0 \times (1 - Se^n)$$

## Data Availability

All data used in this study are publicly available and referenced above.

## Code Availability

The simplified model is available as an Excel spreadsheet, for use under open source license GPLv3 (39).